\documentclass[usenatbib,useAMS]{mn2e}
\usepackage{epsfig,psfig,graphicx,epstopdf,float,color}
\usepackage{subfigure}
\usepackage{mybibdefs}
\usepackage{multirow}
\usepackage{widetext}

\newcommand{\MWA}{\mathrm{MWA}}
\newcommand{\solM}{\mathrm{M_{\odot}}}
\newcommand{\Mpc}{\mbox{Mpc}}

\newcommand{\km}{\mbox{km}}
\newcommand{\s}{\mbox{s}}
\newcommand{\h}{h}

\usepackage{graphicx}
\usepackage{amssymb}
\usepackage{epstopdf}

\title[Does stellar mass assembly history vary with environment?]{Does stellar mass assembly history vary with environment?}  
\author[Hoyle et al.]{Ben  Hoyle$^{1,2}$, Raul Jimenez$^{1,3}$, Licia Verde$^{1,3}$\\$^1$University of Barcelona (UB-IEEC), Marti i Franques 1, Barcelona, 08024 Spain.\\$^2$ICE \&Consejo Superior de Investigaciones Cientificas, Serrano 117, Madrid,
28006, Spain..\\$^3$ICREA \& Institute of Sciences of the Cosmos (ICC), University of Barcelona, Barcelona 08024, Spain. \\\\
{\tt E-mail: benhoyle1212@icc.ub.edu}
 }

\begin{document}
\date{Accepted ----. Received ----; in original form ----.}
\pagerange{\pageref{firstpage}--\pageref{lastpage}} \pubyear{2010}
\maketitle
\label{firstpage}
\begin{abstract}
Using the publicly available VESPA database of SDSS Data Release 7 spectra, we calculate the stellar Mass Weighted Age (hereafter MWA) as a function of local galaxy density and dark matter halo mass. We compare our results with semi-analytic models from the public Millennium Simulation. We find that the stellar MWA has a large scatter which is inherent in the data and consistent with that seen in semi-analytic models. The stellar MWA is consistent with being independent (to first order) with local galaxy density, which is also seen in  semi-analytic models. 

As a function of increasing dark matter halo mass (using the SDSS New York Value Added Group catalogues), we find that the average stellar MWA for member 
galaxies increases, which is again found in semi-analytic models. Furthermore we use public dark matter Mass Accretion History (MAH) code calibrated on simulations, to calculate the dark matter Mass Weighted Age as a function of dark matter halo mass. In agreement with earlier analyses, we find that the stellar MWA and the dark matter MWA are anti correlated for large mass halos, i.e, dark matter accretion does not seem to be the primary factor in determining when stellar mass was compiled. This effect can be described by down-sizing.

 \end{abstract}

\section{introduction}
The spectra of galaxies encodes information about the histories of the component stellar populations, dust, and star formation. Various tools have been developed to extract this information \citep[e.g.,][]{Heavens:1999am,Tojeiro:2007wt} and previous works have compared the resulting extracted information with both extrinsic and intrinsic galaxy properties. These approaches rely on the assumption that the evolution of the stellar populations are well understood and that the current modeling of stellar population is accurate.

The MOPED \citep{Heavens:1999am} routine implements the general process of reforming a complex dataset (e.g., a galaxy spectra) into a set of parameters (e.g., star formation rate, metallicity) and parameter combinations, assuming uncorrelated noise, such that the data compression is loss less \citep[see also][]{2010arXiv1010.5907G}. \cite{Mateus:2008wf} used MOPED-derived stellar masses and luminosities of SDSS Data Release 3 \citep[][]{SDSSDR3} galaxies to build marked correlations functions, and compared with marked correlations of semi analytic models of galaxies in the Millennium Simulation.

More recently, \cite{Ferreras:2010wn} applied a Principal Component Analysis technique, to decompose the spectra of low redshift ($z<0.1$) SDSS early-type galaxies  into two quantities: the average stellar age for the galaxy, assuming metal-rich stellar populations, and the fraction of recent star formation. They found little dependence (and a large scatter) of recent star formation and average stellar age on host halo mass, but find a stronger correlation on local processes, e.g., galactic velocity dispersion. 

An easily accessible, robust code, is the VErsatile SPectral Analysis\footnote{http://www-wfau.roe.ac.uk/vespa/} \citep[hereafter VESPA, see][for more details]{Tojeiro:2007wt,Tojeiro:2009kk} package, which recovers star formation and metallicity histories of the galactic spectra using synthetic stellar population models. The software recovers histories in adaptive age bins according to the signal-to-noise of the galaxy spectrum on a case by case basis and addresses the age-metallicity relation. Two popular synthetic stellar population models are included in the VESPA output, those of \citet[][hereafter BC03]{2003MNRAS.344.1000B}, and the \cite{Maraston:2004em} \& \citet[][]{Maraston:2008nn}  hereafter M05, which differ in their respective resolutions, and the use of empirical libraries to model the thermally pulsating asymptotic giant branch. Furthermore VESPA corrects for galactic extinction using the dust  maps of \cite{dustmaps}, and fits for the dust in each galaxy using either a one or two parameter dust model.

VESPA has been used by, e.g., \cite{Tojeiro:2009kk} to compare the BC03 and M05 models, in particular, the recovered total stellar mass today (finds good agreement) and the mass averaged metallicity (finding less agreement). \cite{Tojeiro:2010up} used VESPA to model the color evolution of high signal-to-noise SDSS Data Release 7 Luminous Red Galaxies as a function of color, luminosity and redshift.

In this paper we compare the results of VESPA, in particular the typical time scale of stellar mass assembly, the stellar Mass Weighted Age (hereafter MWA), with similar results from N-body simulations and semi analytic models. In particular we address the following questions:

\begin{enumerate}
  \item Does the stellar MWA depend on the local density?
  \item Are similar  trends seen in semi analytic models?  
  \end{enumerate}
To this end, we examine the correlation of local galaxy density with the VESPA calculated stellar MWA, and compare with the stellar MWA of the semi analytic models in the Millennium Simulation  \citep{Springel:2005nw,BoylanKolchin:2009nc}.

\begin{enumerate}
\setcounter{enumi}{+2}
  \item Is the stellar MWA correlated with the mass of the dark matter halo that the galaxy inhabits?
\end{enumerate}
In this case, we match the galaxies used in VESPA with those from the New York Value Added Group catalogue \citep[][]{Blanton:2004aa,Yang:2007yr} and explore the average stellar MWA of member galaxies as a function of dark matter halo mass. 

\begin{enumerate}
\setcounter{enumi}{+3}
  \item Is the stellar MWA correlated with the dark matter MWA?
 \end{enumerate}
We address this by comparing the results of the above analysis with the results of Mass Accretion History code \citep{Zhao:2008wd}, rebinned in terms of VESPA time bins for direct comparison to the observations.

The layout of the paper is as follows: in \S\ref{data} we describe the observational and simulated data and the process of reconstructing the stellar  MWA of a galaxy. We continue in \S\ref{method} by detailing our measurement of local galaxy density and dark matter halo mass, for both the observed and semi analytic samples, and present the results in \S\ref{results}. We conclude and discuss in \S\ref{conclusions}. Throughout the paper we employ a flat $\Lambda$CDM cosmology with ($h,\, \Omega_{\Lambda},\,\Omega_{m},\,\sigma_8$) given by ($0.7 \,\km\, \s^{-1}\,\Mpc,0.7,0.3,0.8$)

\section{Data}
\label{data}
In this paper we compare observational data with semi analytic models from simulations, and dark matter mass accretion code calibrated on dark matter simulations, and we describe all data sets below. We also include the Structure Query Language (hereafter SQL) code to retrieve the data sets, and present the charateristics of the final data sets in Table \ref{data_table}.

%
 \begin{table*}
\begin{center}
  \begin{tabular}{l l l l} 
  \multicolumn{4}{| c |}{VESPA runs on SDSS D7 galaxy spectra}\\ \hline
Description   & SP Model \& {\tt runID} & Redshift range& Number of galaxies \\ \hline
Main Galaxy Sample  (MGS)  & BC03 $2$ & $0.05\le z \le0.20$ & $594326$ \\
LRG Sample (LRG) & BC03  $6$ & $0.35\le z \le0.53$ & $25334$ \\
 Main Galaxy Sample   (MGS) &M05  $4$ &  $0.05\le z \le0.20$ & $622749$  \\
LRG Sample (LRG) & M05  $8$ & $0.35\le z \le0.53$ & $75423$  \\ \hline
  \multicolumn{4}{| c |}{Semi Analytic galaxies from the Millennium Simulation  DeLucia2006a table.}\\ \hline
 Redshift & Number of galaxies& Number of parent DM Halos \\ \hline
$z=0.06$ & $21155$ & $7344$\\ 
 $z=0.12$ & $22175$ & $8272$\\ 
 $z=0.17$ & $21818$ & $8230$\\ 
 $z=0.36$ & $138$ & $62$\\
$z=0.41$ & $403$ & $138$\\
 $z=0.51$ & $479$ & $154$\\ \hline
  \end{tabular}
\caption{\label{data_table}A description of the data samples used in this work. We show the acroymn used throughout this paper, the ``runID" from the VESPA database applied to the SDSS DR7 spectra, the redshift cuts, and number of galaxies which have greater than $4$ recovered VESPA age bins. We also show the galaxy data drawn from semi analytic models retrieved from the Millennium Simulation, the redshift of output simulation, the number of galaxies and the number of parent halos.}
\end{center}
\end{table*}

\subsection{Observed Galaxy Data}
\label{data_obs}
The  $6\times10^5$ spectroscopically selected galaxies in this work were drawn from the Sloan Digital Sky Survey Data Release 7 \citep[][and references therein, hereafter SDSS DR7]{York:2000gk,SDSSDR7}. Galaxy targets were selected from SDSS imaging and include ``Main Galaxy Sample" \citep[][hereafter MGS]{Strauss:2002dj}, and ``Luminous Red Galaxies" \citep[][hereafter LRG]{Eisenstein:2001cq}.

Galaxy spectra were reprocessed using VESPA and we followed \cite{Tojeiro:2009kk} in adopting the two parameter dust model. We note that our results are insensitive to the choice of dust models but are more sensitive (to within $20\%$) to the choice of stellar population models, which we discuss in later sections.

We obtained the galaxy data using the following query run on the VESPA SQL web interface\footnote{http://www-wfau.roe.ac.uk/vespa/SQL\_form.html}; modifying the values of \textbf{runid} to obtain different dust models, stellar population models, and galaxy populations. We show the size of the galaxy samples obtained in Table \ref{data_table} and describe the main parts of the query later in this section.

\begin{scriptsize}
\begin{verbatim}
SELECT table3.indexP,table3.AgeMass/table3.TotMass AS MassWeightedAge, 
table3.AgeZ/table3.TotZ AS ZWeightedAge,
table3.AgeSFR/table3.TotSFR AS SFRWeightedAge, 
table3.TotSFR, table3.TotZ, table3.TotMass,
l.specObjID,n.redshift,d.dustval,d.runid, d.dustid, 
n.m_stellar AS TotalCurrentStellarMass,n.nbins,
s.dered_g AS g,s.modelMagerr_g,s.dered_r AS r, s.modelMagerr_r,
s.dered_i AS i, s.modelMagerr_i, s.dered_z AS z, s.z AS spec_z, 
s.zErr AS errspec_z, s.modelMagerr_z,  
s.ra AS gal_ra, s.dec AS gal_dec,
s.MJD, s.FIBERID, s.PLATE, s.TILE, z.absMag_r

FROM (
    SELECT indexP, sum(mass*meanAge) AS AgeMass, sum(mass) AS TotMass,
    sum(Z) AS TotZ, sum(Z*meanAge) AS AgeZ, sum(SFR) AS TotSFR,
    sum(SFR*meanAge) AS AgeSFR

      FROM
         ( SELECT  b.indexp, b.mass AS mass,
         (bID.ageStart+bID.ageEnd)/2 AS meanAge, b.Z, b.SFR

             FROM binprop AS b
             
             JOIN  binID AS bID  ON
             ( b.binID = bID.binID AND b.runid = 8)

             JOIN lookupTable AS l ON ( b.indexP = l.indexP)

         ) AS table2 group by indexp
       
     ) AS table3
     
JOIN lookupTable AS l ON ( table3.indexP = l.indexP )
JOIN galprop n ON (table3.indexP=n.indexP AND n.runid=8)
JOIN dustprop d ON (table3.indexP=d.indexP AND d.runid=8 AND dustid=2)
JOIN bestDR7..SpecPhotoAll  AS s ON (s.specObjID=l.specObjID )
JOIN bestDR7..photoz  AS z ON (z.ObjID=s.ObjID )
\end{verbatim}
\end{scriptsize}

In this work, we are concerned with one possible output of the VESPA database, namely the stellar Mass Weighted Age ($\MWA$). It is calculated, as in the above SQL query, as the sum over VESPA age bins $i$, of the mass $M_i$, formed in the bin, multiplied by the central bin age $t_i$, divided by the total stellar mass formed in the galaxy i.e.
\begin{eqnarray}
\label{awm1}  \MWA & =  & \frac{1}{ M_{tot}} \sum_i t_iM_i \:.
\end{eqnarray}

The adaptive binning inherent to VESPA adopts wide age bins for spectra with low signal-to-noise, leading to an unnatural peak in the MWA distribution at $7$ Gyrs \cite[which corresponds to the central value of the largest time bin, see Fig. 1 of][]{Tojeiro:2009kk}. To remove these unphysical solutions we further imposed that the number of recovered VESPA bins is greater than $4$, when determining averaged stellar MWA.  We use the full data sets when calculating the local galaxy density. In Table \ref{data_table} we present the final redshift ranges and the number of galaxies for both the MGS and the LRG samples using the different stellar population models.

\subsection{Simulated Galaxy Data}
The simulated data were drawn from the Millennium Simulation, which used a modified version of the collisionless dark matter and gas dynamical cosmological simulation code GADGET2 \citep{2001NewA....6...79S,2005MNRAS.364.1105S} with $2160^3$  dark matter particles,  each with mass $8.6\times 10^8\, h^{-1} \solM$, in a volume of side  $500\, \h^{-1} \Mpc$. The simulation is able to recover dark matter halos small enough to host luminous galaxies brighter than $0.1L_{\star}$.

The dark matter halos were populated with galaxies according to the semi analytic models of \cite{DeLucia:2005yk} and \cite{DeLucia:2006vua}.  The models are novel in that they continue to follow dark matter halos until they are tidally disrupted by parent halos, and allow central galaxies  to replenish their supply of cold gas, unlike satellite galaxies. They contain prescriptions for gas infall onto halos, mixing between hot and cold gaseous states and cooling to form stars, and allow for cooling flows and their suppression by central galaxy Active Galactic Nuclei activity \citep[see][for specific details]{Springel:2005nw,2006MNRAS.365...11C,2004MNRAS.349.1101D}.

Both the galaxy properties and host dark-matter halo properties are stored in the ``DeLucia2006a" table found on the public data access website\footnote{http://www.g-vo.org/Millennium}. We obtained information about local galaxy density, within a box of side $3\h^{-1}\Mpc$ centered on each galaxy, and the stellar mass weighted age, using the below query.
 
\begin{scriptsize}
\begin{verbatim}
SELECT d.galaxyID, d.haloID, d.stellarMass, d.snapnum, s.redshift, 
s.lookBackTime, d1.galaxyID as galaxyid2 , d.massWeightedAge, 
d1.massWeightedAge AS massWeightedAgeGal, d.centralMvir ,sdss.r_sdss

FROM DeLucia2006a d JOIN Snapshots s ON s.snapnum=d.snapnum 
AND d.snapnum=60  JOIN DeLucia2006a d1 ON d.snapnum=d1.snapnum 
AND d1.snapnum=60 AND ABS(d.x-d1.x) <1.5 AND abs(d.y-d1.y) <1.5 
AND ABS(d.z-d1.z) <1.5 AND  ABS(d.centralMvir - Msval) < 50  
JOIN DeLucia2006a_SDSS2MASS AS sdss on d1.galaxyid=sdss.galaxyid
\end{verbatim}
\end{scriptsize}

We called this query multiple times by looping over increasing values of  ``Msval" to obtained all halos with central halo dark matter mass within $|$ d.centralMvir - Msval $|$$<50$. This was performed to circumvent timeout errors. We chose to obtain data from the simulations with snap numbers $60,58,56,51,50,48$ which roughly correspond to the centers of the redshift bins used in the analysis which follows.

\section{Methodology}
\label{method}
In this section we describe the construction of the local galaxy densities and the assignment of dark matter halo masses to the observational and simulated semi analytic data.  We note that here the density is only dependent on cosmology through the volume element, which changes by less than $6\%$ ($9\%$) from  $0.2<z<0.5$  in a flat fiducial $\Lambda$CDM when changing the cosmological parameter $\Omega_m$ from $0.3$ to $0.25$ (or $h$ from $0.7$ to $0.72$).

\subsection{Observational local galaxy density }
The magnitude limits of the SDSS imply that we cannot directly compare the number density of all observed galaxies at different redshifts. We therefore determined the maximum volume out to which each galaxy could have been observed, and weighted the density by this factor. This is the standard $V_{max}$ correction \citep{1968ApJ...151..393S} and the maximum volume was calculated by spline fitting to the faint end of the band absolute magnitude $M_r$ and redshift $z$ distributions of the MGS and LRGs separately. The equation of the spline fit is given by $z= 6.12 +  0.91M_r +    0.04 M_r^2 +   7\times10^{-4}M_r^3$ for the LRGs and $z=-262.22 -31.52 M_r  -1.26 M_r^2  -0.016M_r^3$ for the MGS. We additionally removed galaxies fainter than the spline fit. 

We calculated the local density in a cylindrical cell of radius $r_0=2.25\h^{-1}\Mpc$ (i.e. diameter $4.5\,\h^{-1}\,\Mpc$) and length $l_0=4.5\h^{-1}\Mpc$  and additionally weighted each galaxy by its $V_{max}$ correction. We also examined the effect on our results of changes in the cylinder volume, by modifying both the radius and length to $r_1=1.5$ \& $l_1=3\,h^{-1}\,\Mpc$ and $r_2=3$ \& $l_2=6\,h^{-1}\,\Mpc$. We note that the dependence of stellar MWA with density  for the LRGs using both stellar population models, are indifferent to the changes in the cylinder volumes. For the BC03 model applied to the MGS, the stellar MWA at lower densities also remain unchanged, but at higher densities there is an increase of $\sim 0.5$ Gyrs for  radii increases from $r_0$ to $r_2$,  and the MWA decreases by $\sim 1$ Gyrs for radii decreases from $r_0$ to $r_1$. For the M05 models applied to the MGS, the stellar MWA also remains unchanged at low densities, but at higher densities it decreases by $1.5$ Gyrs for radii changes  from $r_0$ to $r_1$.

Edge and survey holes were accounted for by creating a pixelised mask using the Hierarchical Equal Area isoLatitude Pixelization\footnote{http://healpix.jpl.nasa.gov} \citep[hereafter HEALPix]{Gorski:2004by} routine, applied to a large SDSS DR7 photometric galaxy sample with resolution \textbf{Nside}$=128$, and only retaining pixels if they were completely surrounded by other pixels. At this HEALPix resolution there are many pixels per cylindrical cell. We divided the galaxy distributions into comoving distances slices such that there were three slices per cylinder length.  

Each pixel in each redshift slice, which contained at least one galaxy became the center of a cylinder (which itself spanned the redshift slice in front and behind of the redshift slice in question), within which we calculated the $V_{max}$ weighted density for all galaxies in the cylinder (irrespective of the number of recovered VESPA age bins), and we further divided this by the total number of galaxies within the three redshift slices encompassing the cylinder. We then calculated the average stellar MWA of only the galaxies which resided in the central pixel.

\begin{figure*}
   \centering
     \subfigure[ \label{Vweight1a} BC03 model ]{  \includegraphics[scale=0.4]{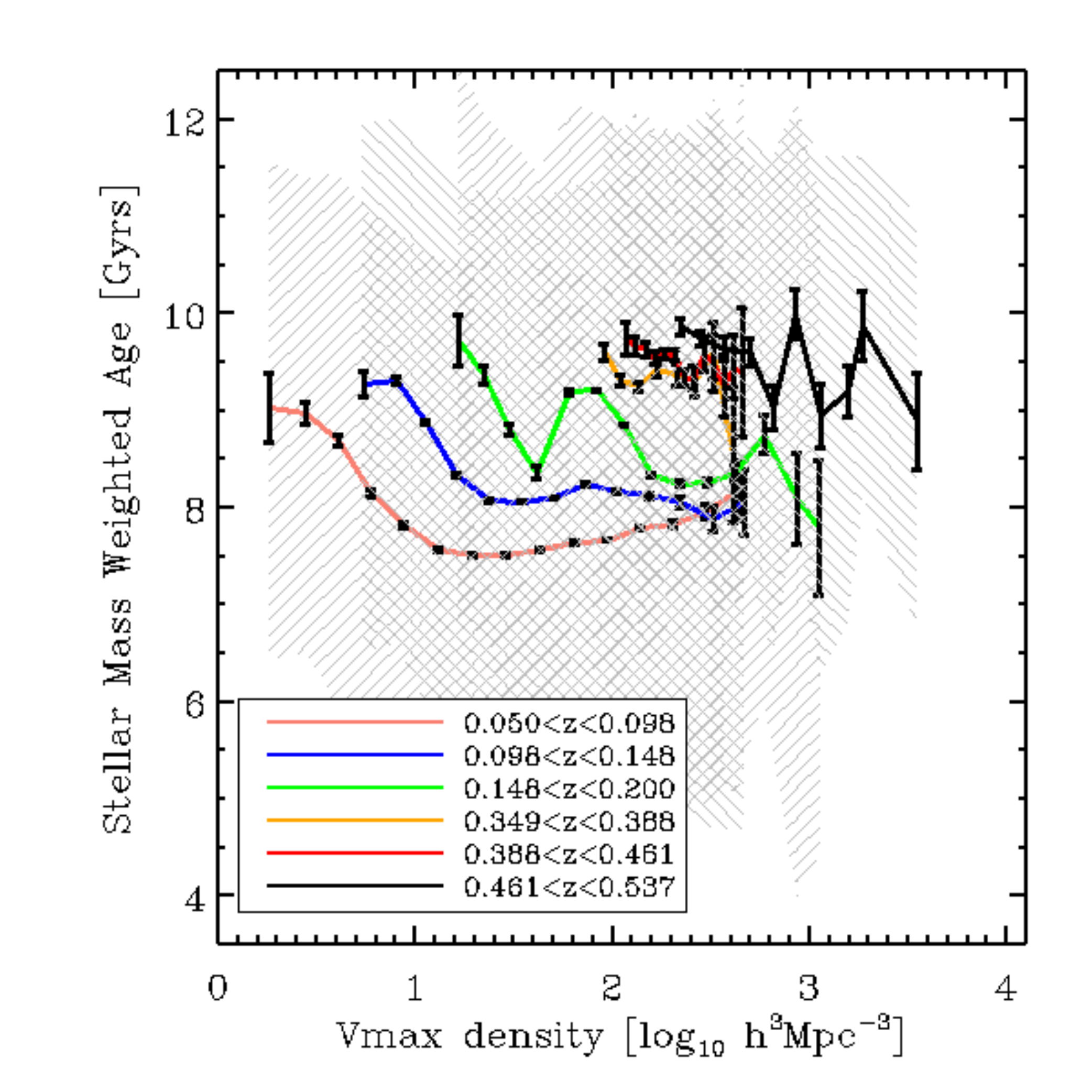} }
      \subfigure[ \label{Vweight1b} M05 model]{  \includegraphics[scale=0.4]{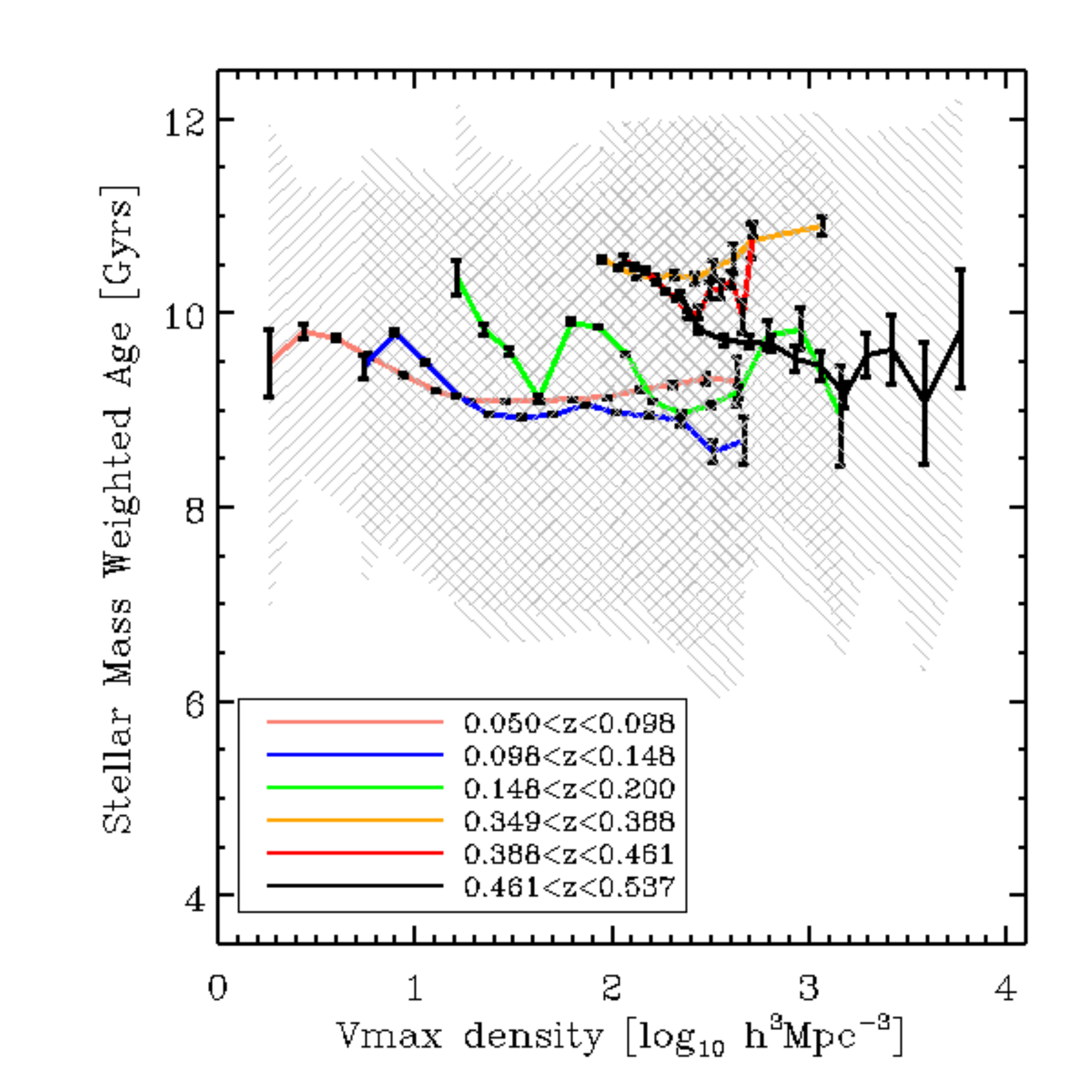} 	}
          \subfigure[ \label{Vweight1c}SA model]{  \includegraphics[scale=0.4]{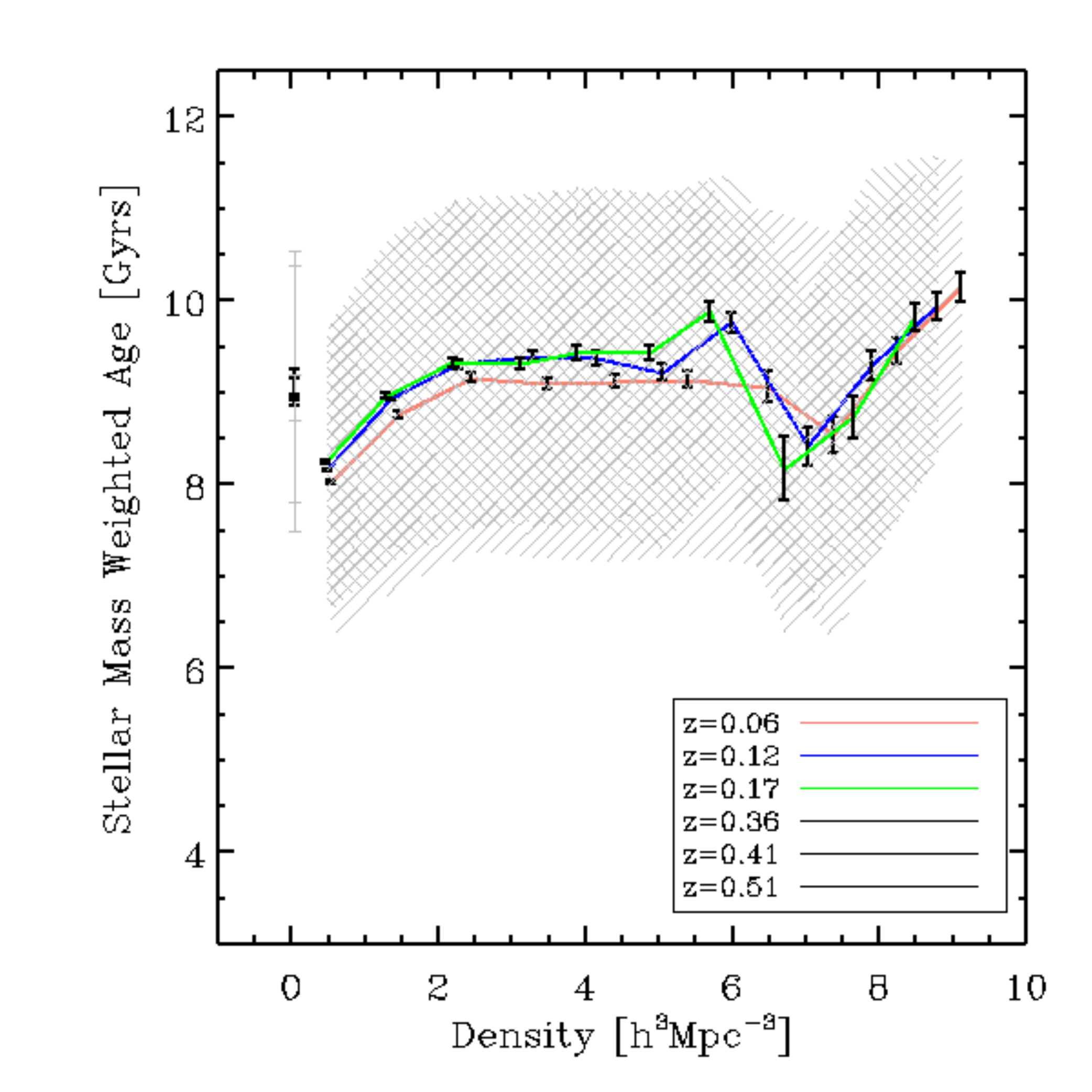}}
   \caption{  \label{Vweight1} The mass weighted age as a function of $V_{max}$ weighted density, for SDSS galaxies. Fig \ref{Vweight1a} shows the mass weighted age reconstructed using the BC03 stellar population models, and Fig \ref{Vweight1b} uses the M05 models. Fig \ref{Vweight1c} shows the mass weighted age of galaxies drawn from the Millenium Simulations by applying semi analytic (SA) models to the dark matter halos. The higher redshift LRGs and lower redshift MGS are each subdivided into three redshift bins, shown by the color lines and black points.  We show the $1\sigma$ standard deviation by the grey region, and the error on the mean with the black error bars.}
\end{figure*}

\begin{figure*}
   \centering
    \subfigure[ \label{AMW_haloMass1a}BC03]{  \includegraphics[scale=0.4]{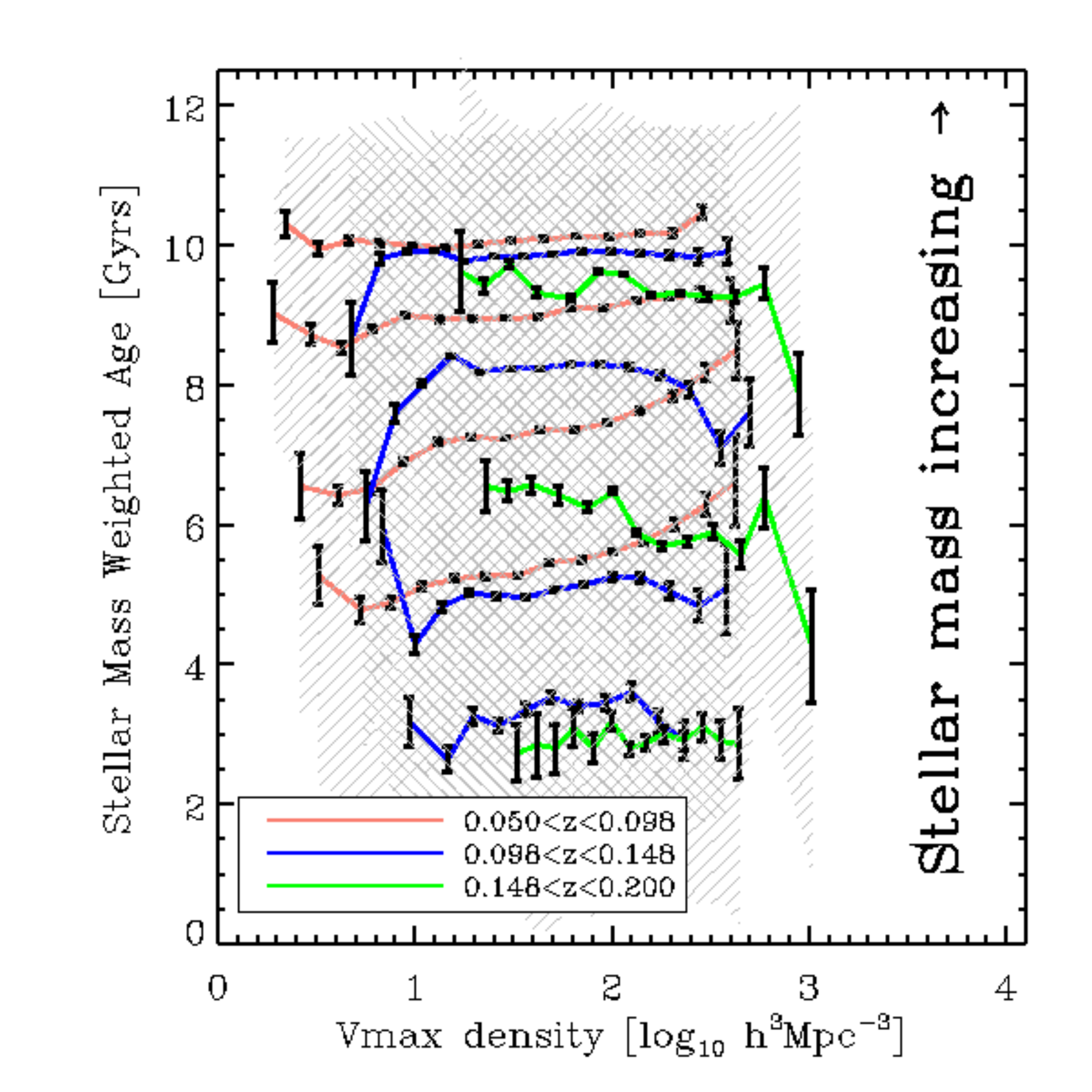} 	}
     \subfigure[ \label{AMW_haloMass2a}M05]{  \includegraphics[scale=0.4]{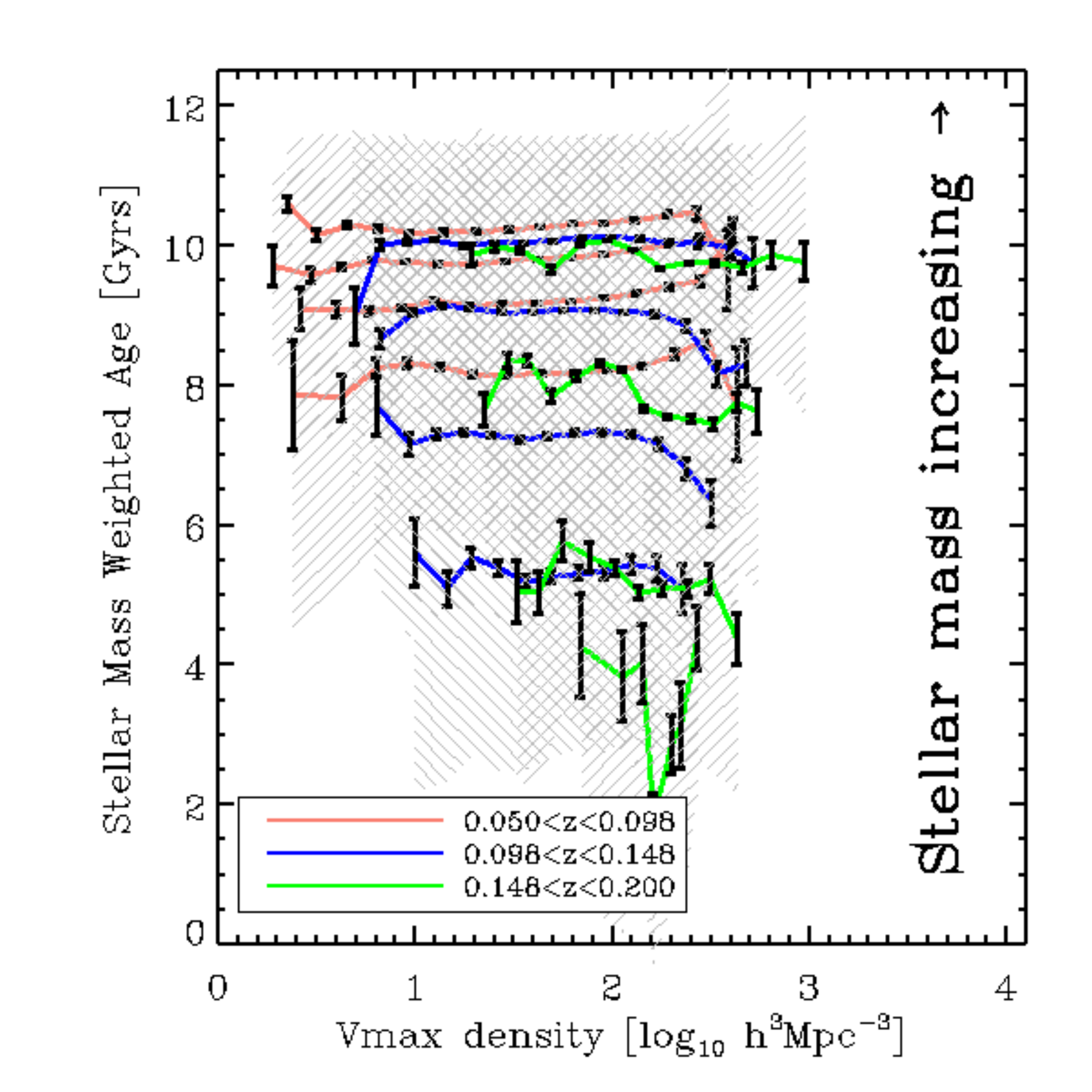} } 
   \caption{  \label{AMW_haloMassa1aa} The stellar MWA and $V_{max}$ weighted density relation of the low redshift MGS, subdivided into bins of total stellar mass for the BC03 model (left panel) and M05 model (right panel). We use the colored lines to distinguish redshift slices, and like colored lines of increasing stellar MWA correspond to bins of increasing total stellar mass. 
The stellar mass $M$ bins are given in units of $\solM$, by the ranges $10.2\le \log_{10} M < 10.4$, $10.4\le \log_{10} M < 10.8$, $10.8\le \log_{10} M < 11.2$ and $11.2\le \log_{10} M <11.6$, which split the sample into bins of roughly equal number. The error on the mean is again showed by the black error bars, and the $1\sigma$ dispersion by the grey hashed region.}
\end{figure*}

\subsection{Observationally-estimated dark matter halo masses}
We matched the SDSS DR7 galaxies in our sample to the group members catalogue \citep{Yang:2007yr} of the New York University Value-Added Galaxy Catalogue \citep[NYU-VAGC][]{Blanton:2004aa}, which was run on the SDSS DR4 \citep[][]{AdelmanMcCarthy:2005se} Main Galaxy Sample. The group catalogue prescribes a dark matter halo mass to galaxy groups and clusters as a function of the total optical luminosity of galaxies above a characteristic luminosity \citep[see][for details]{Yang:2007yr}.  This optical-based dark matter halo mass was found to agree with the ``true'' dark matter halo mass when compared with mock galaxy redshift surveys. For each dark matter halo, we calculated the mean of the stellar MWA of galaxy members.

As we were only able to estimate the dark matter halo masses of the low redshift MGS, we have restricted our comparison of the galaxies from the semi analytic models, to the low redshift snapshots of the Millennium Simulation. We calculated the average stellar MWA of all galaxies in each parent dark matter halo.

\subsection{Local galaxy density and halo masses from simulations}
Using the SDSS data we determined the maximum absolute $r$ band magnitude of galaxies within each redshift slice, and applied the same absolute magnitudes cuts to the galaxies drawn from the semi analytic models.  These cuts were $-16.24$ for the three lower redshift snapshots,  $-22.66$ for the higher redshift snapshots. The magnitude cuts applied to the higher redshift simulation outputs result in a lower number of recovered galaxies and halos. The output of the SQL query returned the number of galaxies  within a box of side $3\,h^{-1}\,\Mpc$ centered on each galaxy.We then simply calculated the number of galaxies in each box which passed the absolute magnitude cuts and divided by the box size. We then recorded the stellar MWA of the galaxy at the center of the box.  We additionally calculated the average stellar MWA of all galaxies (which, again had passed the absolute magnitude cuts) in each parent dark matter halo.

\section{Results}
\label{results}
\subsection{Local galaxy density}
In Fig. \ref{Vweight1} we present the stellar MWA of VESPA galaxies using the BC03 and M05 models, as a function of $V_{max}$ weighted density. We also show the stellar MWA of the semi analytic model galaxies, as a function of local galaxy density at a range of redshifts chosen to coincide with  the observed VESPA galaxies. The $1\sigma$ dispersion of the stellar MWA is indicated using the grey hashed regions, and the error on the mean by the black error bars.  We show the median relations for each redshift by the colored lines.

Both Fig. \ref{Vweight1a} and  \ref{Vweight1b} show that the SDSS MGS sample probe lower local densities than the higher redshift LRG sample, as expected from observations \citep[e.g.,][]{Bamford:2008ra}.  

We find that there is a large spread in stellar MWA as a function of  $V_{max}$-weighted density for both stellar population models with the VESPA data, and also find a similar stellar MWA dispersion for the semi analytic models, see Fig \ref{Vweight1c}, although the density parameter has not been $V_{max}$-corrected because the simulations are not magnitude dependent. Weighting the semi analytic model galaxies by the  $V_{max}$ correction could add extra scatter.

We can try to understand if the large dispersion is actually a feature of the data, if it is due to systematic offsets in the stellar population models, or the noise in the galaxy spectra. Comparing the medians (the colored lines) of the stellar MWA of the two different synthetic stellar population models run on the almost identical input data sets (the SDSS galaxies),  we see that there is typically a $1-2$ Gyr difference between them, which is indicative of a systematic offset. Additionally we can examine the results of \cite{Tojeiro:2010up}, who stacked high signal-to-noise spectra from the SDSS DR7 LRG sample \citep{SDSSDR7}, and compared VESPA outputs (e.g., redshift evolution of metallicity weighted age, MWA) between the BC03 and M05 models, see their Fig 6. They also find a $\gtrsim 20\%$ dispersion between the models, but also find that both models need to be modified to include additional error terms to fit the observed spectra well. These results suggest that the dispersion is at least partly due to the uncertainty in the models and systematic differences, independent of uncertainties in the data. If we now compare the dispersion to that seen in the galaxies drawn from the semi analytic model, we conclude that the remaining scatter is a combination of being intrinsic to the data and due to noise.

\begin{figure*}
   \centering
    \subfigure[ \label{AMW_haloMass1} Dark matter MAH]{  \includegraphics[scale=0.4]{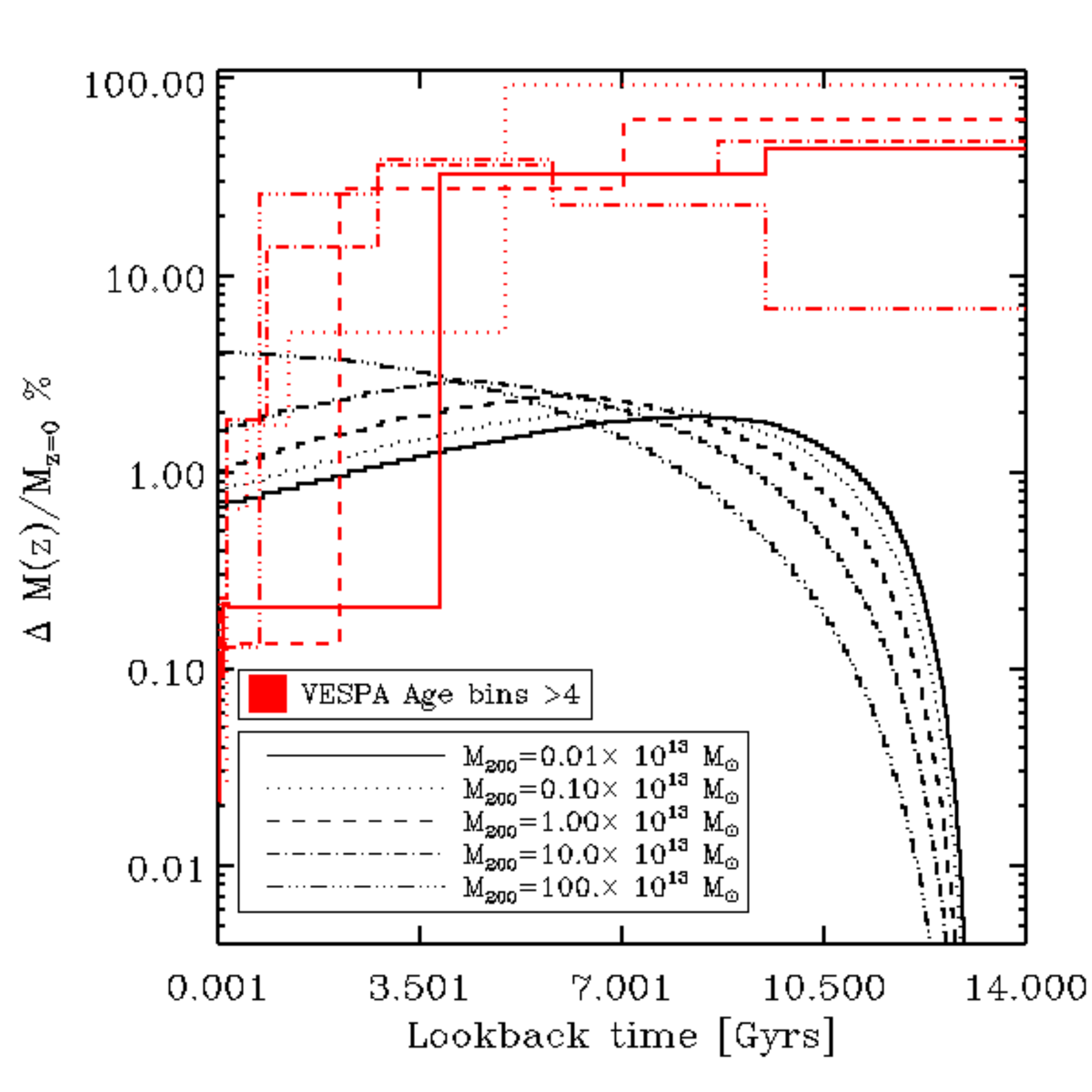} 	}\\
     \subfigure[ \label{AMW_haloMass2} BC03 \& M05 models, with DM MWA]{  \includegraphics[scale=0.4]{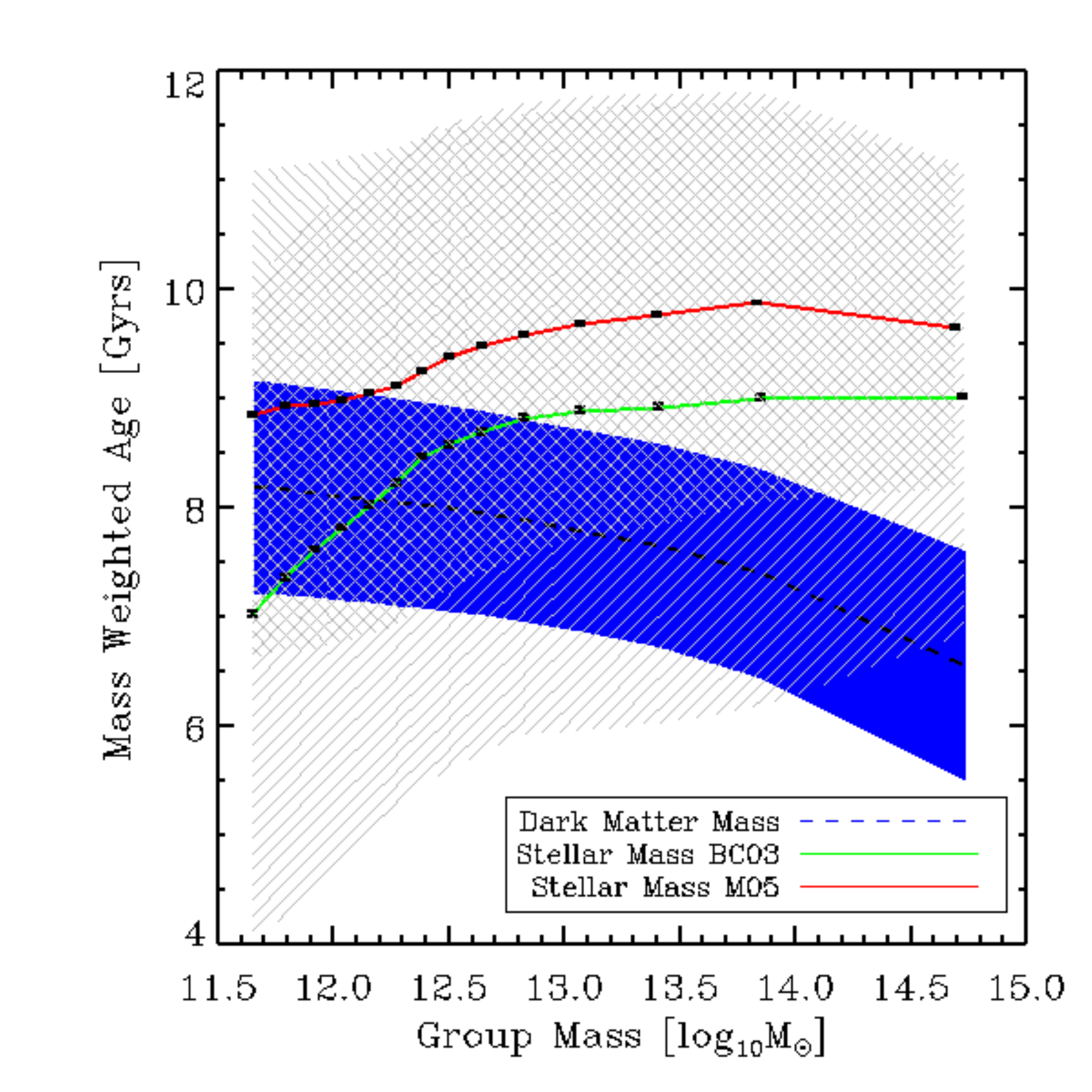} } 
         \subfigure[ \label{AMW_haloMass3} Semi analytic models]{  \includegraphics[scale=0.4]{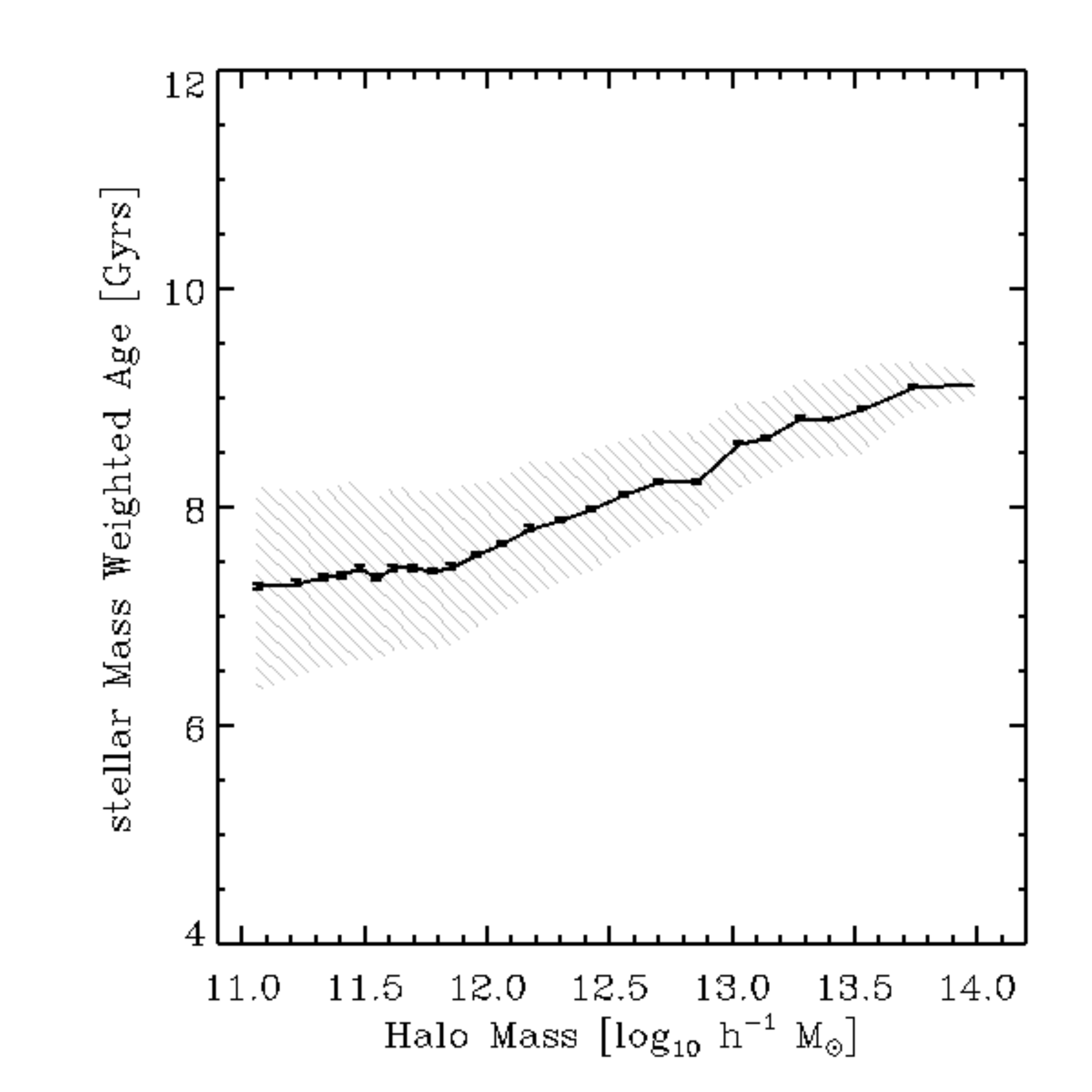} }   
   \caption{  \label{AMW_haloMass} We show the dark matter assembly history, using the MAH code, and the assembly history in age bins reconstructed as could be measured with VESPA in Fig \ref{AMW_haloMass1}.  Fig \ref{AMW_haloMass2} shows the stellar MWA as a function of group halo mass for both the M05 models (red) and BC03 models (green), and the $1\sigma$ dispersion by the grey hashed region, and the error on the mean by the black lines. We also show the dark matter MWA (DM MWA), as calculated by MAH code (black dotted lines) and show the $1\sigma$ dispersion on the dark matter MWA with the blue regions, which are determined by $10^{4}$ random re-samples using the bin ages available to VESPA.  We additionally show, in Fig \ref{AMW_haloMass3}, the stellar MWA as a function of halo mass, for galaxies from the semi analytic models in the Millennium Simulation, and again show the $1\sigma$ dispersion by the grey hashed region, and the error on the mean by the black error bars. }
\end{figure*}

Examining Fig. \ref{Vweight1a} and  \ref{Vweight1b} we see that, as noted above, the two different stellar population models agree to within $\sim20\%$, and that in both cases there is a $\sim1$ Gyr difference between the stellar MWA of LRGs and MGS, which is a consequence of an older stellar population, i.e. there has been comparatively little or no recent star formation in LRGs compared with MGS. We note that a direct comparison of the BC03 and M05 stellar MWA for the $21801$ LRG galaxies with measured stellar MWA and the constraint $Nbins>4$ has a large scatter ($68\% = 3.7$ Gyrs) and a small systematic offset of $0.6$ Gyrs.  A possible explanation of the systematic offset, is that, of the two stellar population models used in VESPA, the M05 had a lower wavelength resolution than BC03. VESPA re-bins the data to the models, so the signal-to-noise is larger in the M05 models, and thus more age bins are recovered. Also there is a decrease in star formation between  $0.1$ and $1$Gyr using BC03, which will mean there will be little stellar mass compiled in the corresponding VESPA age bins, i.e., the stellar MWA will be skewed low.

Continuing to examine Fig. \ref{Vweight1a} and \ref{Vweight1b} we also find that the stellar MWA for all galaxies is typically older than $8 \rightarrow 9.5$ Gyrs in look back time for the MGS, and typically older than $9\rightarrow10$ Gyrs for the LRGs. We note that Fig. \ref{Vweight1c} is divided by redshift, but does not distinguish between LRG and MGS galaxy populations (i.e. they are combined here), but do produce the same trends, i.e. that the largest fraction of stellar assembly occurred before $8$ Gyrs look back time.

For each galaxy sample analysed with the BC03 model, we find that the median stellar MWA of the three increasing redshift slices increases. This suggests that even though the majority of star formation occurred at high redshift, both the LRG and MGS samples continuously have small amounts of stars being assembled, thereby reducing their stellar MWA. This effect is less apparent with M05 model, with only the two upper redshift slices of the MGS obeying this trend.

This weak trend is however consistent with observations of the continual Star Formation Rates (SFR) of both low and high mass galaxies across cosmic time \citep[][although note, that the SFR for high redshift massive galaxies is greater than for high redshift low mass galaxies]{2009ApJ...696..620C,2007ApJ...660L..47N,2007ApJ...660L..43N}.

All panels of Figs. \ref{Vweight1} appear to show no, or little dependence of stellar MWA on local galaxy density to first order. Also note that the observed and simulated trends agree with each other within the dispersion. Furthermore, we see no discernible trends between the different redshift slices of either the MGS or LRGs as a function of density.

A correlation of very local galaxy density \citep[related to projected density of the $5^{th}$ nearest neighbour,][]{2005ApJ...634..833C} and galaxy age estimates following \cite{2005MNRAS.362...41G} have led \cite{2010MNRAS.402.1942C} to find a strong relation between galaxy age and local galaxy density at fixed stellar mass. Our estimates of density are in much larger volumes, and thus these smaller scale correlations are smoothed out.

In Fig. \ref{AMW_haloMassa1aa} we have split the stellar MWA and $V_{max}$-weighted density relation into bins of total stellar mass for both models. We use the colored lines to distinguish redshift slices, and like colored lines of increasing stellar MWA correspond to bins of increasing total stellar mass. The stellar mass $M$ bins are given  in units of $\solM$, by the ranges $10.2\le \log_{10} M < 10.4$, $10.4\le \log_{10} M < 10.8$, $10.8\le \log_{10} M < 11.2$ and $11.2\le \log_{10} M <11.6$, which split the sample into bins of roughly equal galaxy number. The error on the mean is again showed by the black error bars, and the $1\sigma$ dispersion by the grey hashed region.  We have only shown the mass splitting of the lower redshift MGS to make the figures more readable. There is very little stellar MWA splitting between the bins of equal mass for each redshift slice of the LRG sample. 

The stellar MWA is consistent with being independent on local galaxy density, independent of total stellar mass bin, redshift slice or stellar population model, within the $1\sigma$ spread of the data.  However, we do find that the stellar MWA of lower-mass galaxies is consistently lower than more massive galaxies, across all redshift slices and independent of the stellar population model.  This suggests that the lower the mass of a galaxy, the more stellar mass is compiled at late times, i.e., low mass galaxies are younger than high mass galaxies.  This result has been observed before, e.g., \cite{2005MNRAS.356..495J,2007MNRAS.378.1550P} found that the star formation rates of massive galaxies peaked earlier than less massive galaxies.

\subsection{Dark Matter Halo Mass}
We now compare both the stellar MWA (for low-redshift galaxies matched to the DR4 NYVAC group catalogue) and the dark matter MWA, to the dark matter host halo mass. To do so we used the publicly available code of \cite{Zhao:2008wd} which calculates the Mass Accretion Histories (hereafter MAH)  of dark matter halos back through cosmic time, by assuming that the progenitor of a halo is always the most massive of all progenitors. The fitting formula was calibrated on suites of N-body dark matter simulations \citep[see][for details]{Zhao:2008wd} and was found to be accurate to within $\lesssim 5\%$ and universal. Results may be defined in terms of $M_{200}$, which enabled direct comparison with the halo masses of the NY VAC group catalogue. 

In Fig. \ref{AMW_haloMass1} we show, as a function of halo mass, the percentage of accreted dark matter onto the dark matter halos per time interval, relative to the halo mass at  $z=0$. The black lines show the time steps as outputted by the MAH code, and the red lines show different realizations of the larger time steps corresponding to a random selection of $>4$ VESPA age bins. The VESPA age bins were chosen by imposing the constraint that each random sample of bin selections be consecutive and complete across the full VESPA time range. This age bin selection was performed to most closely match the conditions of the VESPA data; recall that each included galaxy spectrum has a high signal-to-noise, such that VESPA could determine stellar histories in $>4$ different age bins. The different line styles correspond to different mass halos at $z=0$.

In Fig. \ref{AMW_haloMass2} we show the average stellar MWA of halo member galaxies as a function the dark matter group halo mass, and distinguish between the M05 models (red) and BC03 models (green). We show the $1\sigma$ dispersion by the grey hashed region, and the error on the mean by the black error bars. We again note that the spread of data is large, and that the models agree to within $\sim1.5$ Gyrs.  For both models we see a slight increase in the stellar MWA for increasing  dark matter halo mass, which suggests that more massive groups and clusters formed the majority of their stars earlier in the Universe's history than less massive groups, or similarly, that stellar mass was still being assembled in lower mass groups more recently. We additionally show the dark matter MWA by the black dotted line. The blue contours correspond to  $10^{4}$ dark matter MWA measurements drawn from random samplings of the $30$ VESPA age bins, with the constraints (as introduced above), that we randomly choose more than $4$ VESPA bins, and that the bins form a full and continuous coverage of VESPA ages.  We see that the dark matter MWA becomes lower as the group mass increases, and that the $1\sigma$ simulated dispersion is constant as a function of group mass.

The stellar and dark matter MWA of the smaller systems have similar values, but this may not be a physical correlation because of the anti-correlation of the stellar and dark matter MWA at higher masses. This leads us to infer, that while the stellar MWA is correlated with the overall mass of the dark matter halo, is it not correlated with the accretion of dark matter onto the halo.

Fig \ref{AMW_haloMass3} shows the stellar MWA as a function of dark matter halo mass, as measured from galaxies using semi analytic models in the Millennium Simulation. We note that the masses probed by the simulations is lower than those from the NYVAC, but we do see a clear increase in the stellar MWA  with increasing dark matter halo mass, in agreement with the VESPA data, and again opposite to the trends seen in the dark matter MWA.

\section{Conclusions}
\label{conclusions}

We used the results of VESPA \citep{Tojeiro:2007wt,Tojeiro:2009kk} which analyzed $10^6$ SDSS DR7 \citep{SDSSDR7} galaxy spectra, to calculate the stellar Mass Weighted Age (MWA). To remove peculiarities in the data, we chose to only use high signal-to-noise spectra which allowed the measurement of stellar histories in greater than $4$ VESPA age bins. We then calculated the local galaxy density of the all the SDSS galaxies in cylinders of radius $2.25\,\h^{-1}\,\Mpc$ and length $4.5\h^{-1}\,\Mpc$, and applied the standard $V_{max}$ correction to account for the magnitude limits of the SDSS.  We also cross-matched the VESPA galaxies with the New York Value Added Group Catalog \citep{Yang:2007yr}, which includes estimates of the parent dark matter halo mass. 

Using the public code of \cite{Zhao:2008wd},  we determined the dark matter MWA as a function of halo mass. We additionally obtained the stellar MWA, local galaxy density, and dark matter halo masses of galaxies from within the Millennium Simulation \citep{Springel:2005nw}, populated according to the semi analytic models of \cite{DeLucia:2005yk}.

We found a large scatter in the recovered stellar MWA, which we determined to be partially due ($\sim 20\%$) to the differences between the BC03 \citep[][]{2003MNRAS.344.1000B} and M05 \citep{Maraston:2008nn} stellar populations models, by comparing the medians of our distributions, and also by examining the results of \cite{Tojeiro:2010up}, who compared the differences between models when applied to stacked LRG spectra with high signal-to-noise. A large dispersion is also observed in the stellar MWA of the semi-analytic models, suggesting that the dispersion is also inherent to the data.  We found that the recovered stellar MWA using the different stellar population models agree to within $1.5$Gyrs and that the stellar MWA of most ($60\%$ to $90\%$ depending on stellar population model and galaxy sample) galaxies was older than $8$ Gyrs, independent of redshift, local galaxy density, dark matter halo mass or galaxy type. This is expected from the star formation history of the Universe, which peaks at around $z=1-2$, corresponding to $8-10$ Gyrs lookback time \citep[e.g., see][]{2004Natur.428..625H}.

We now return to the questions posed in the introduction;
\begin{enumerate}
  \item Does the stellar MWA depend on the local density?
  \item Are similar  trends seen in semi-analytic models?  
  \end{enumerate}
To first order, we found that stellar MWA does not appear to be related to local galaxy density in either the observed or simulated data. We found similar dispersions in the observed and simulated data and apparent flatness across local galaxy density \citep[although see][]{2010MNRAS.402.1942C}.

\begin{enumerate}
\setcounter{enumi}{+2}
  \item Is the stellar MWA correlated with the mass of the Dark Matter Halo that the galaxy inhabits?
\end{enumerate}
We did find a correlation of older stellar MWA with increasing dark matter halo mass in the observed galaxy sample, independent of stellar population model, which was also seen in the semi-analytic models, albeit for a smaller range in dark matter halo mass.

\begin{enumerate}
\setcounter{enumi}{+3}
  \item Is the stellar MWA correlated with the dark matter MWA?
 \end{enumerate}
We found that the dark matter MWA became anti correlated with the stellar MWA as the mass of dark matter halos increases.  This is an observation of a ``downsizing'' effect  \citep[][]{1996AJ....112..839C}  which describes the idea that the dark matter halo mass at which star formation is highest, shifts to lower masses at later times \citep[see e.g., Fig 8 \& 9 of ][]{2009ApJ...696..620C,2005MNRAS.356..495J}. Another meaning of ``downsizing'' is that more massive galaxies compiled more of their stars at higher redshift (and in a shorter time scale) than less massive galaxies as seen in Fig. \ref{AMW_haloMassa1aa} and e.g., \cite{2005ApJ...621..673T,2007ApJ...669..947J}.  These results have been seen previously by measuring the Specific Star Formation Rates (the amount of star formation per solar mass) of sets of similar mass galaxies over a range of redshifts. 

Here, the downsizing effect is observed by the decrease in the stellar MWA  as a function of decreasing stellar and dark matter halo mass, using VESPA stellar history reconstructions of the galaxy spectra. These conclusions are in agreement with other measures of ``downsizing''  using MOPED to reconstruct the Star Formation Rate (SFR) histories of SDSS DR3 galaxies \citep[][found the SFR of massive galaxies peaked earlier than less massive galaxies]{2007MNRAS.378.1550P}.  

We find VESPA is well suited for reconstructing stellar histories using the included synthetic stellar population models, and agrees well with semi-analytical models drawn from larger simulations.

\section*{Acknowledgments} 
\label{ack}
The authors would like to thank Rita Tojeiro for VESPA database support and useful suggestions, and Aday Robaina for insightful discussions and feedback which have improved the manuscript.  BH would like to thank the Theory Division, Department of Physical Sciences, University of Helsinki for office space, and acknowledges funding from grant number FP7-PEOPLE- 2007- 4-3-IRG n 20218, LV is supported by FP7-PEOPLE-2007-4-3-IRG n. 202182, FP7-IDEAS-Phys.LSS 
240117; LV and RJ are supported by  MICINN grant AYA2008-03531
Funding
for the SDSS and SDSS-II has been provided by the Alfred
P. Sloan Foundation, the Participating Institutions, the
National Science Foundation, the U.S. Department of
Energy, the National Aeronautics and Space Administration,
the Japanese Monbukagakusho, the Max Planck
Society, and the Higher Education Funding Council for
England. The SDSS Web Site is http://www.sdss.org/.

\bibliographystyle{mn2e}
\bibliography{awm}
\end{document}